\documentclass[prb,twocolumn,showpacs,amsmath,amssymb]{revtex4}
\usepackage{graphicx}
\usepackage{dcolumn}
\usepackage{bm}
\begin{document}
\newcommand{\be}{\begin{eqnarray}}
\newcommand{\ee}{\end{eqnarray}}
\title{Fine structure of phonon replicas in a tunnel spectrum of a GaAs quantum well.}
\author{V. G. Krishtop\footnote{krishtop@iptm.ru}}
\affiliation{Institute of Microelectronics Technology, Russian Academy of Sciences, Chernogolovka, Moscow region, 142432 Russia}
\affiliation{Moscow Institute of Physics and Technology, Dolgoprudny, Moscow region, 141700 Russia}

\author{ V. G. Popov }
\affiliation{Institute of Microelectronics Technology, Russian Academy of Sciences, Chernogolovka, Moscow region, 142432 Russia}
\affiliation{Moscow Institute of Physics and Technology, Dolgoprudny, Moscow region, 141700 Russia}
\author{M. Henini}
\affiliation{School of Physics and Astronomy of Nottingham University, UK}
\author{Yu. Krupko}
\affiliation{LNCMI, UPR 3228, CNRS-INSA-UJF-UPS, BP166, 38042 Grenoble cedex 9, France}
\author{J.-C. Portal}
\affiliation{LNCMI, UPR 3228, CNRS-INSA-UJF-UPS, BP166, 38042 Grenoble cedex 9, France}
\affiliation{Institut Universitaire de France and Institut National des Sciences Applique'es, Toulouse, 31077 cedex 4, France}
\date{\today}

\begin{abstract}
A fine structure of phonon replicas in the current-voltage characteristic of a resonant-tunneling diode has been investigated experimentally. A detailed study of the diode I-V curves in magnetic fields of different orientations has allowed  to determine the origin of the features in the fine structure. The voltage positions of the features are  shown to coincide with calculated that in the frame of two models: LO-phonon assisted tunneling and resonant tunneling of polarons.
\end{abstract}
\maketitle
\section{Introduction}
Resonant tunneling spectroscopy is a powerful method to investigate numerous effects in low-dimensional structures, such as interactions of electrons with each other\cite{Pseudogap2010}$^,$\cite{effects2006}, phonons\cite{LB89lofonons}, polarons\cite{Polarons2010} and plasmons\cite{Plasmonassisted}. Resonant-tunneling diodes (RTD) have a huge potential for practical applications\cite{japanese2} because the diodes are the fastest solid-state devices operating up to 1 THz \cite{Sugiyama}. The latest realization of RTD on the graphene\cite{kostja2013} suggests a wide area for future applications.
Resonant character of tunneling is very important for this because it provides a narrow current peak when the difference between subband energies in emitter and in QW is zero and satellite current peaks when the energy difference is equal to energies of LO-phonons\cite{LB89lofonons}$^,$\cite{japanese2}, cyclotron energy\cite{Smoliner91}.
To explain the phonon replica a model of LO-phonon assisted tunneling (LOPAT) is commonly used [9]. In this model the additional current is caused by an electron tunneling with the emission of LO-phonons. In this case the phonon replicas should be spaced from a resonance current peak by voltages proportional to the energies of the LO-phonons, as follows:
\begin{equation}\label{EqLOPAT}
eU_{LO1,2} = eU_{peak} + \alpha \varepsilon_{1,2}
\end{equation}
where $e$ is the carriers charge, $U_{LO1,2}$ are the phonon-replicas positions in LOPAT model, $\alpha$ is a leverage factor, $U_{peak}$ is a main resonant peak position, $\varepsilon_1$,$\varepsilon_2$ are energies of the LO-phonons. In the case of the GaAs the energy of the LO-phonon is $\varepsilon_1 = 36$~meV, for AlGaAs the energy is $\varepsilon_2 = 53$~meV.
In the Ref \onlinecite{Polarons2010} another mechanism for the phonon-replicas origin had been proposed, it is a model of  resonant tunneling of polarons (RTP). In this model resonant tunneling of electrons occurs from the 2D-accumulated layer in emitter to specific 2D polaron subbands formed in the 2D spectrum of the quantum well (QW) between tunnel barrier of the RTD. In this case the phonon replicas should be observed at lower voltage than in LOPAT model as follows.
\begin{equation}\label{EqRTP}
eU_{p1,2} = eU_{peak} + \alpha (\varepsilon_{1,2} - \varepsilon_F)
\end{equation}
where $U_{p1,2}$ are the phonon replicas positions in PRT model, $\varepsilon_F$ is a Fermi energy in the emitter accumulation quantum well.
An in-plane applied magnetic field broadens the main resonance and shifts the main resonance to higher voltages\cite{Smoliner91}. A method of the magneto-tunneling spectroscopy used the in-plane magnetic field\cite{mts87}$^,$\cite{mts1}$^,$\cite{mts2}$^,$\cite{mts3}. The main equation of the magneto-tunneling spectroscopy for the voltage position of the resonant current peak is following:
\begin{equation}\label{Eq0}
eU_{peak} = \alpha E_{QW}(eB \Delta z)
\end{equation}
where $E_{QW}(p)$ is the 2D dispersion of electrons in the QW, $\Delta z = z_1 - z_2$ is a tunneling length, where $z_1$ and $z_2$ are electron averaged positions in the emitter and QW states, and $B$ is a magnetic field. In accordance with this equation, the magnetic dependence of $U_{peak}$ reproduces the 2D dispersion of the electrons in the direction perpendicular to the magnetic field \cite{Polarons2010}. The crucial point is to determine the leverage factor because it is an aspect ratio between voltage and energy difference of the subband levels. Usually its value is found from data obtained in a perpendicular magnetic field.
In the  magnetic field applied perpendicular to the layers plane new peaks appear in the I-V curves. This peaks originated from electron tunneling between the Landau levels with different indexes in QWs. The voltage position difference $\Delta U_n$ between the new features $U_n$ and the resonance $U_{peak}$ \cite{Smoliner91} or its phonon replicas $U_{LO1,2}$ corresponds to energy difference between the Landau levels with different indexes as follows \cite{leadbeater89}:
\be\label{cycl}
e\Delta U_n = U_n - U_{peak, LO1,2} = \alpha n \hbar \omega_c
\ee
where $n$ is the index difference and $\omega_c$ is the cyclotron frequency. Since the cyclotron energy is well-known value and doesn't depends upon the QW profile, one can easily to determine $\alpha$ from this Equation (\ref{cycl}).
In the paper we investigate the fine structure of the phonon replica of the RTD made of asymmetric double-barrier heterostructure (see Section \ref{Samp}). The replica is investigated at voltage when the charge isn't build-up in the QW. This provides a relatively constant value of the $\alpha$ and one can extract the subband-energy difference from the bias voltage precisely. In Section \ref{Perpen} the features of the I-V curves are considered in the magnetic field directed perpendicular to the layer plane. In Section \ref{Inplane} we have determined leverage factor from the I-V curves data measured in the in-plane magnetic field. In Section \ref{Dis} we compare the data from different Sections and with theoretical models. Finally we conclude the results in Section \ref{Conc}.
\section{\label{Samp}Sample and its spectrum}
\begin{table}[h!]
\caption{\label{tab1}Layer sequence of the heterostructure under investigation.}
\renewcommand{\arraystretch}{1.4}
\begin{center}
\begin{tabular}{ccccc}
\hline
\hline
\rule{0cm}{0.3cm}
$num$ & $layer$ & $material$ & $doping$ & $thickness$ \\
\hline
$1$ & top contact & GaAs & $2\times10^{18}cm^{-3}$ & $2\mu m$ \\
$2$ & & GaAs & $1\times10^{17}cm^{-3}$ & $50 nm$\\
$3$ & spacer & GaAs & $1\times10^{16} cm^{-3}$ & $50 nm$\\
$4$ & & GaAs &undoped& $3,3 nm$\\
$5$ & thin barrier & $Al_{0.4}Ga_{0.6}As$ & undoped & $8.3 nm$ \\
$6$ & well & GaAs & undoped & $5.8 nm$ \\
$7$ & thick barrier & $Al_{0.4}Ga_{0.6}As$ & undoped & $11.1 nm$ \\
$8$ & & GaAs & undoped& $3.3 nm$\\
$9$ & spacer & GaAs & $1\times10^{16} sm^{-3}$ & $50 nm$ \\
$10$ & & GaAs & $1\times10^{17} cm^{-3}$ & $50 nm$ \\
$11$ & & GaAs & $2\times10^{18}$ & $500 nm$ \\
\hline
\multicolumn{5}{m{70mm}}{\centering{substrate}}\\
\hline
\hline
\end{tabular}
\end{center}
\end{table}
\begin{figure}[h]
\includegraphics[width=0.9\linewidth]{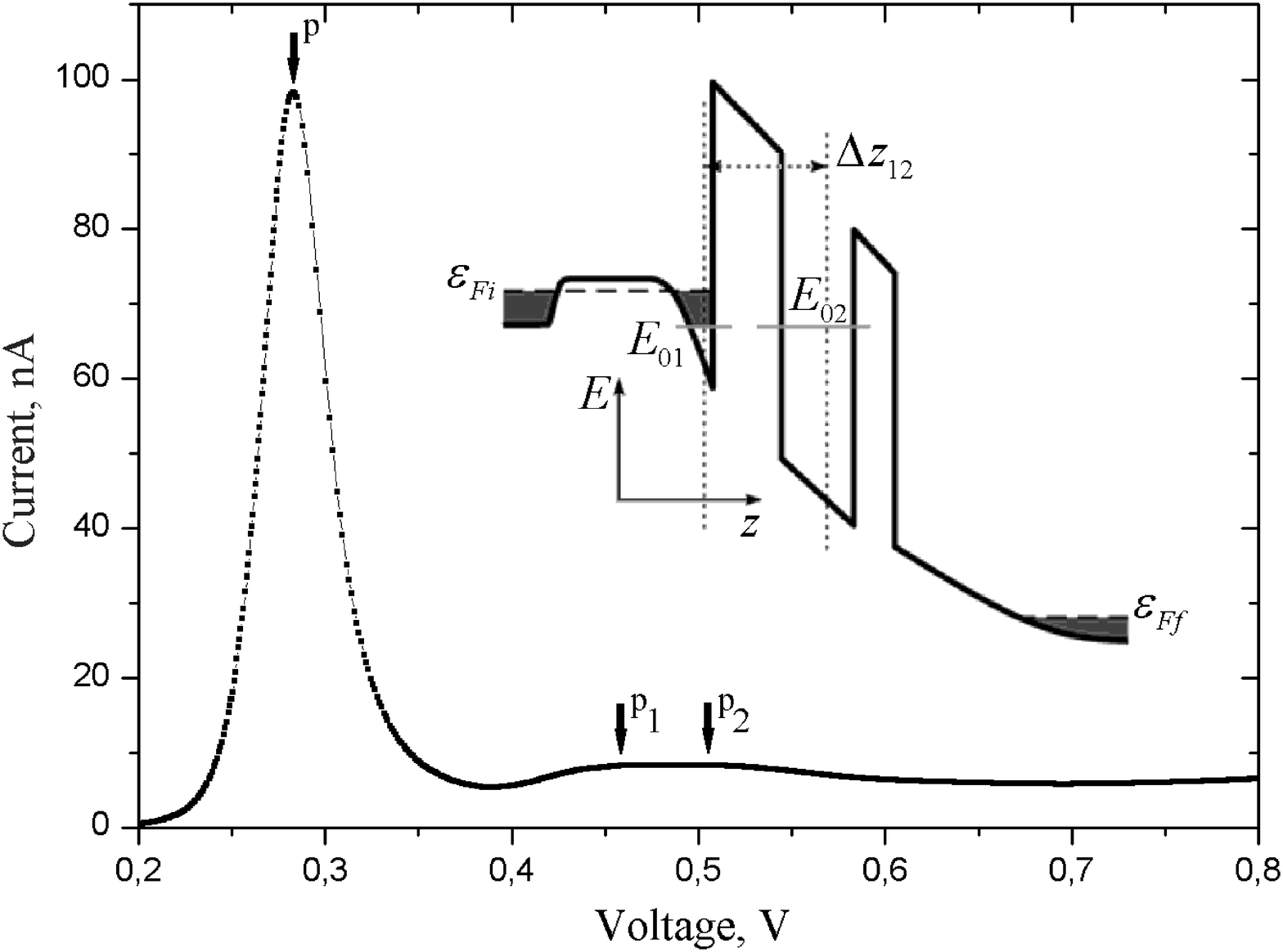}
\caption{\label{fig1} I-V curve of the RTD. Arrows shows positions of current maxima. In the inset the RTD conduction-band-bottom profile is shown with subband levels at an applied bias voltage $V_b$.}
\end{figure}
\begin{figure}[h!]
\includegraphics[width=0.9\linewidth]{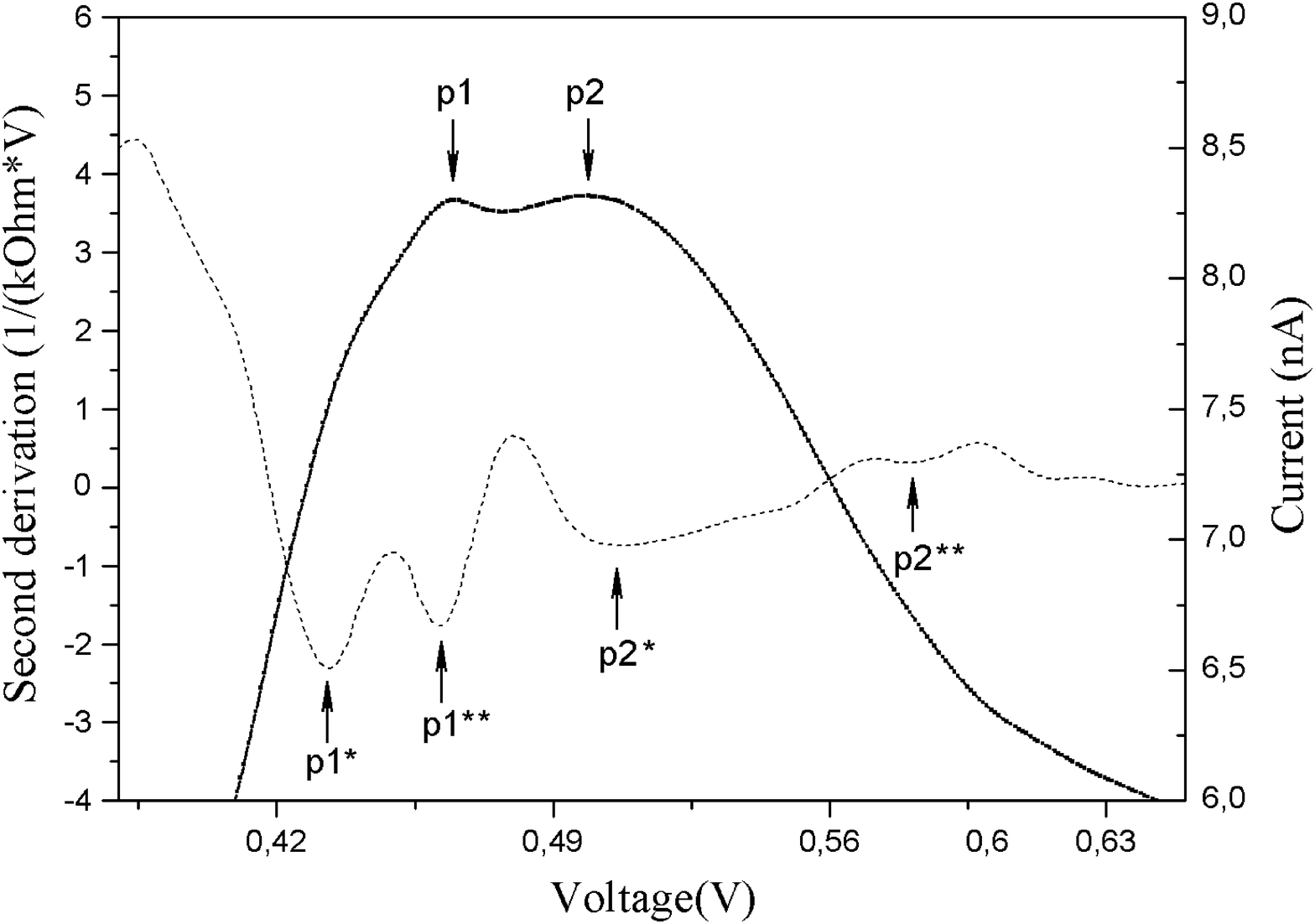}
\caption{\label{fig10} I-V curve of RTD (solid) and the second derivative of the I-V curve (dotted) of RTD in phonon replica voltage range. Arrows shows positions of current maxima on I-V curve, and minimums on second derivative curve. Minimums of the second derivative curve corresponds to features of the I-V curve.}
\end{figure}
Tunnel diode is made of an asymmetric double-barrier heterostructure. Sequence of the structure layers is shown in Table \ref{tab1}.
Briefly speaking the GaAs QW layer has a thickness of 5.6 nm and has been grown between Al$_{0.4}$Ga$_{0.6}$As barrier layers of thicknesses of 11 nm and 8.3 nm. The barriers are separated from heavily doped layers of GaAs by thin undoped layers 3.3 nm and low-doped spacers layers of GaAs of the thickness of 50 nm. The diodes were made of the heterostructure by conventional technique of wet etching, photolithography and contact annealing.

In Fig.\ref{fig1} a current-voltage characteristic of the diode is shown at bias polarity when a charge build-up in the QW is minimal that is thicker barrier is on the emitter side (see insert (a) in the Fig.\ref{fig1}). All measurements were conducted at the liquid He$^3$ temperature $T = 0.3$ K.
The main current peak $p$ corresponds to the main resonance when the ground subband levels are aligned, i.e., $E_{01}(U_{peak}) = E_{02}(U_{peak})$. A split current peak (phonon replica) can be as usual described with LOPAT.
In this model two additional peaks are associated with the emission of the phonons with energies $\varepsilon_1$ and $\varepsilon_2$. The second origin of the peaks could be RTP. The peaks in the RTP model also associated with these phonons, but RTP peaks are shifted from the LOPAT peaks on the voltage corresponded to the Fermi energy in the 2D accumulated layer in the emitter. To study the phonon replicas in details we have considered the second current derivative (see dashed curve in Fig. \ref{fig2}). In this case the derivative minima correspond to the current maxima. One can see in Figure \ref{fig2} that there are four derivative minima $p_1^*$, $p_2^*$, $p_1^{**}$ and $p_2^{**}$. To explain they one need to know $\varepsilon_F$ and $\alpha$ and also to be sure that these parameters are constant in the voltage range from the current peak $p$ up to derivative minimum $p_2^{**}$. Usually these parameters are found from the I-V curves in the magnetic field.
\section{\label{Perpen}Perpendicular magnetic field}
\begin{figure}[h]
\includegraphics[width=0.9\linewidth]{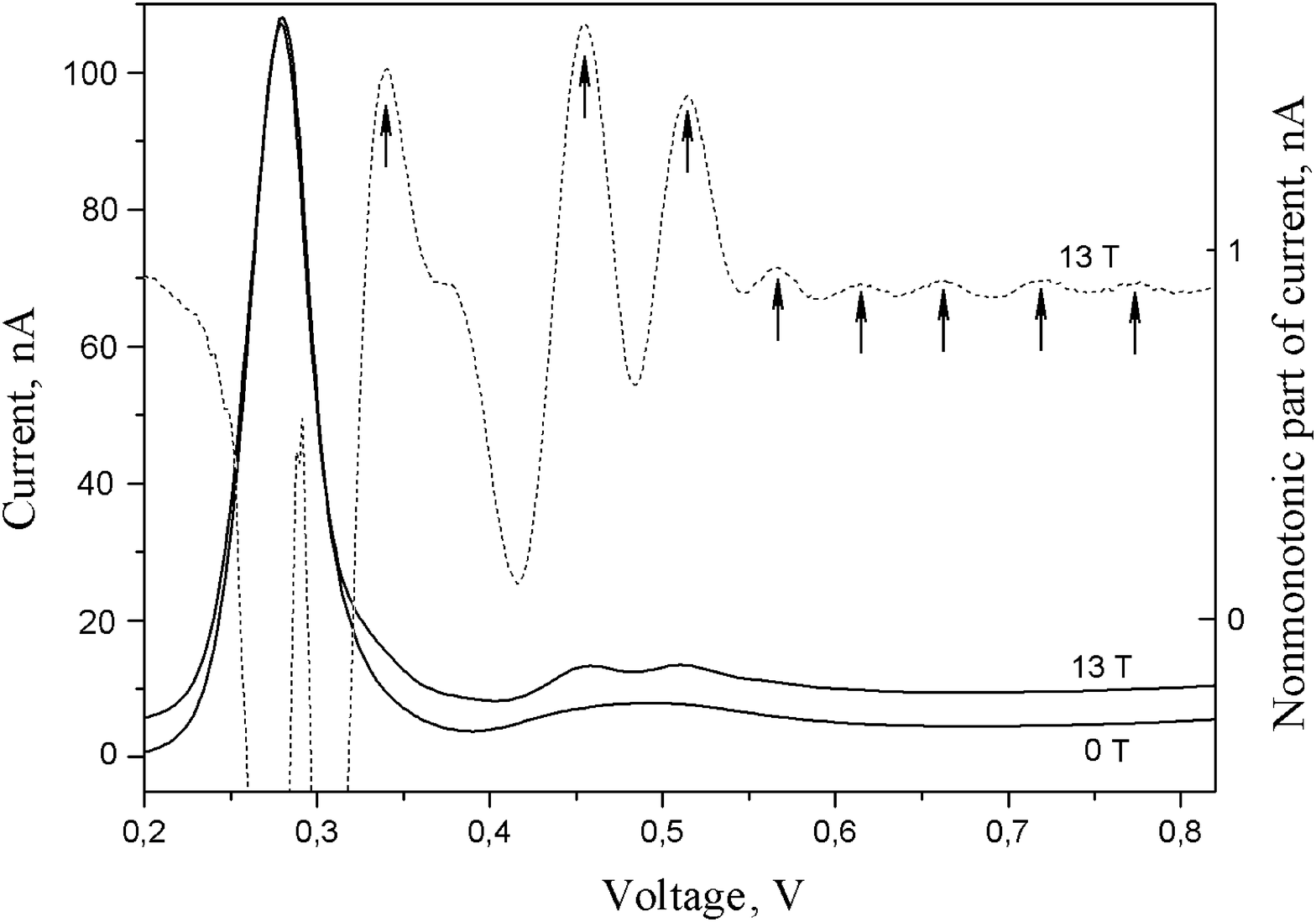}
\caption{\label{fig6} Current-voltage characteristics  in zero and at 13 T magnetic fields are shown by solid curves. The curves are shifted relative to one another. Dotted curve corresponds to a nonmonotonic part of the current at 13 T magnetic field. Arrows indicates the clearly resolved  peaks.}
\end{figure}
Additional features have been observed in the I-V curves  in magnetic field directed perpendicular the QW plane (see Fig.\ref{fig6}). To clarify these features the background current has been substracted from the each I-V curves, i.e. current in the zero magnetic field is subtracted. This procedure allows to confidently identify the peaks in the current. The result of the substraction, i.e. nonmonotonic part of the current, also is shown in Fig.\ref{fig6}.
\begin{figure}[h]
\includegraphics[width=0.9\linewidth]{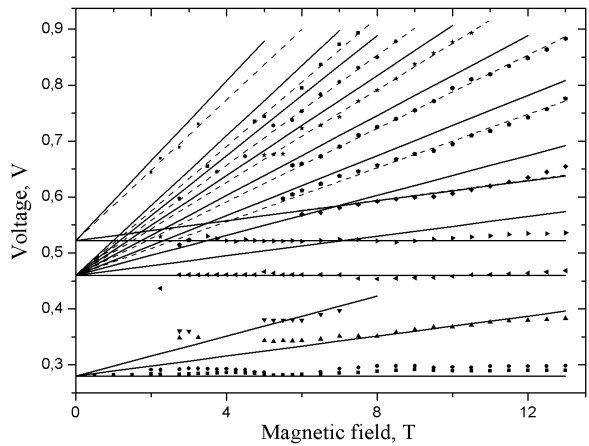}
\caption{\label{fig8} Voltage positions $U_n$ of the additional peaks on the perpendicular magnetic field. Experimental values are shown by symbols, solid lines formed a fan-diagram for $\alpha = 5.1$, dashed lines are the best-fitted  lines starting from the positions of current peaks on the I-V curve at zero field.}
\end{figure}
Voltage positions of the additional peaks $U_n$ in the current due to tunneling between the Landau levels is shown in Fig. \ref{fig8}. Dashed lines in Fig. \ref{fig8} are obtained by fitting the experimental data by straight lines starting from the positions $U_{peak}$, $U_{p1}$, $U_{p2}$ on the I-V curve at zero field. We have determined slopes of each line and  calculated $\alpha$, applying Equation (\ref{cycl}) to every pair of the adjacent best-fitted lines for three groups of lines or fan-diagrams. The calculated values are within the range $4.9 < \alpha <5.3$ with the average value of 5.1. This indicates good stability of the leverage factor both in the peak current area and in the area of phonon replicas. Solid lines in Fig. \ref{fig8} are forming a fan-diagram for $\alpha = 5.1$. It can be seen that dashed lines do not coincide exactly with solid lines. There are many reasons for this because the Equation (\ref{cycl}) does not take in to account such effects as magnetic dependence of the polaron energy and QW potential profile. Nevertheless, good coincidence of the best-fitted lines with experimental data, and voltage and field independence of their slope differences  let us to say that the value of alpha is defined correctly.

\begin{figure}[h]
\includegraphics[width=0.9\linewidth]{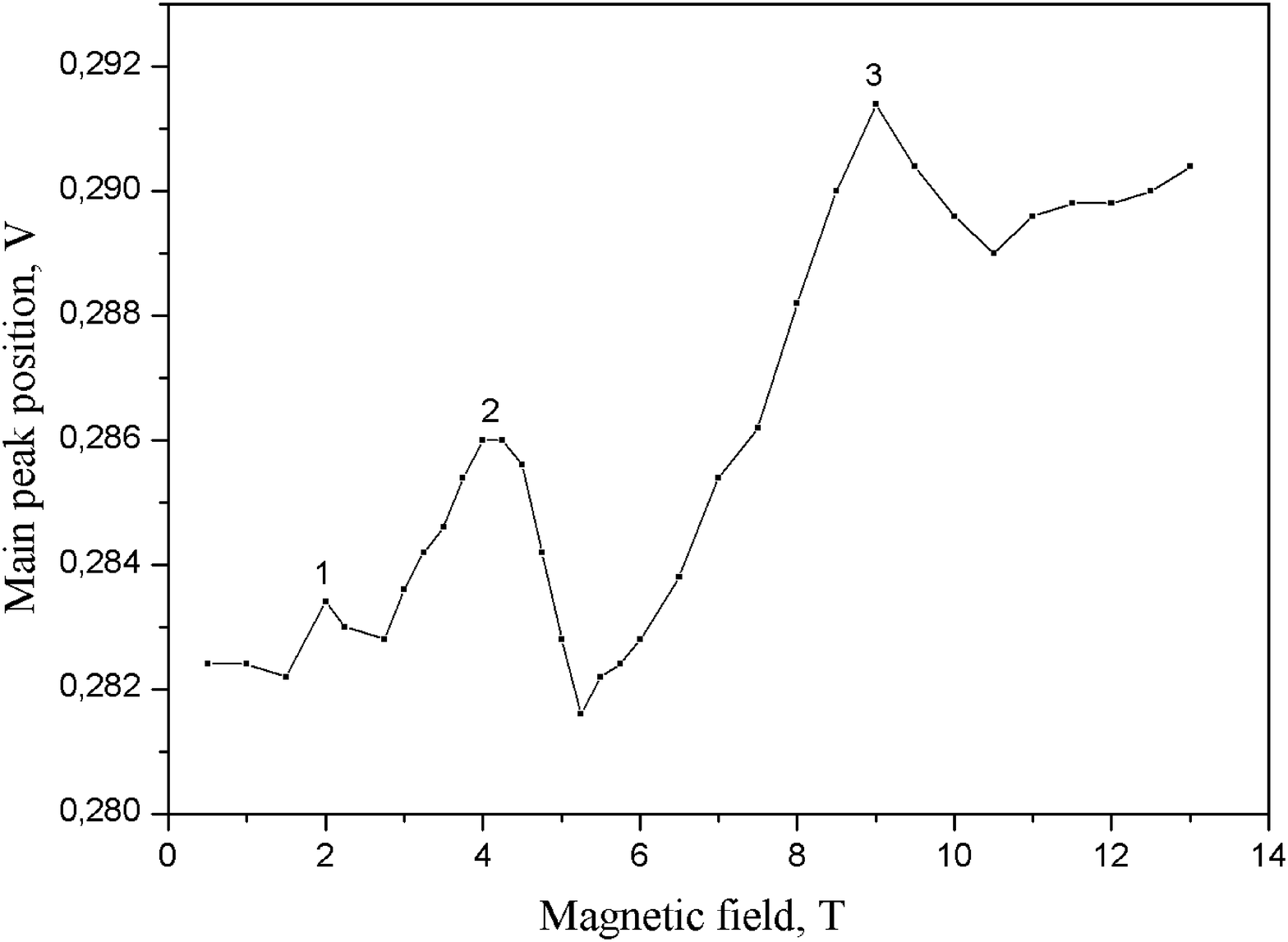}
\caption{\label{fig9} Voltage positions $U_{peak}$ in perpendicular magnetic field.
}
\end{figure}
The $\varepsilon _F$ can be estimated from the magnetic dependence of the resonance voltage $U_{peak}$. Magnetic dependence of $U_{peak}$ is shown in Fig. \ref{fig9}. Similar nonmonotonic dependence of the $U_{peak}$ had been theoretically investigated in Ref. \onlinecite{Popovtheory} and experimentally observed at 2D-2D tunneling in single-barrier structures in Ref. \onlinecite{Popov2}. This dependence appears due to the Landau level pinning in accumulation 2D layer in the emitter. Partially filled Landau level (LL) is pinned to Fermi level in the emitter due to the transfer of electrons out the 2D layer to the heavily doped emitter contact layer. In this case the magnetic field increase causes the potential change that decreases the energy of the subband level $E_{01}$. As a result the resonant voltage $U_{peak}$ increases. When the LL is emptied the electrons transfer back to the 2D layer and next LL jumps to the Fermi level due to the $ E_{01}$ increase. This causes $U_{peak}$ decrease. In this case the LL filling factor is integer and one can determine the average value of the electron concentration and $\varepsilon _F$. As one can see in Figure \ref{fig9} the last $U_{peak}$ decrease at a field 9T. This corresponds to the LL filling factor $\nu$ equals 1. Here the LL are supposed to be spin split. The LL spin splitting can be confirmed if one compares the distance between the peaks 1 and 2 and the distance between peaks 2 and 3 in units of inverse field differs almost twice. Considering $\nu = 1$ at $B = 9$ T we estimate the Fermi energy from the expression $\varepsilon_F=\frac{1}{2}\hbar\omega_c$, and $\varepsilon_F=8~meV$.
The parameter $\alpha$ can also be determined from behavior of the resonance peak in the in-plane magnetic field directed parallel to the layers. It is important to compare these values, because they can be dependent on current and thus complicate the tunnel spectroscopy or features identification.
\section{\label{Inplane} In-plane magnetic field}
\begin{figure}[h]
\includegraphics[width=0.9\linewidth]{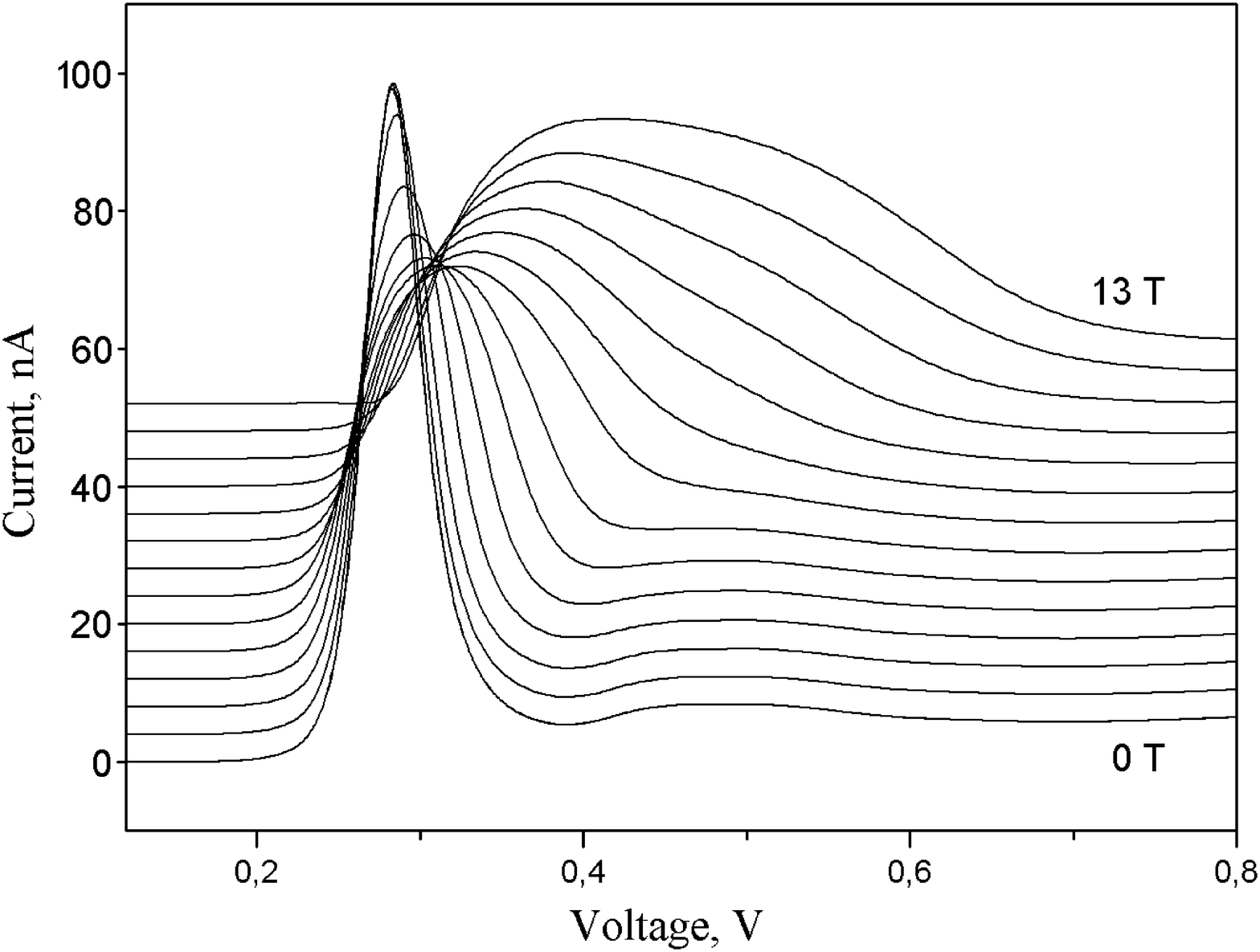}
\caption{\label{fig2} Current-voltage characteristics in the in-plane magnetic fields up to 13 T. The curves are shifted along vertical axis  with step 4 nm that corresponds to the step in magnetic field of 1~T.
}
\end{figure}
In the magnetic field applied parallel to the heterostructure layers the current peak is broadened and shifted to higher voltages (see Fig.\ref{fig2}). This shift and broadening are followings of the electron generalized momentum conservation in tunneling and, as it was shown in Reference \onlinecite{Polarons2010}, the current peak should shifts on a voltage value $U_{peak}$ which is determined from Eq. \ref{Eq0} as follows:
\be \label{Eq1}
\frac{eU_{peak}}{\alpha} = \frac{1}{2m^*}(e \Delta z B)^2
\ee
where is $m^*$ is the effective mass of carriers.
\begin{figure}[h]
\includegraphics[width=0.9\linewidth]{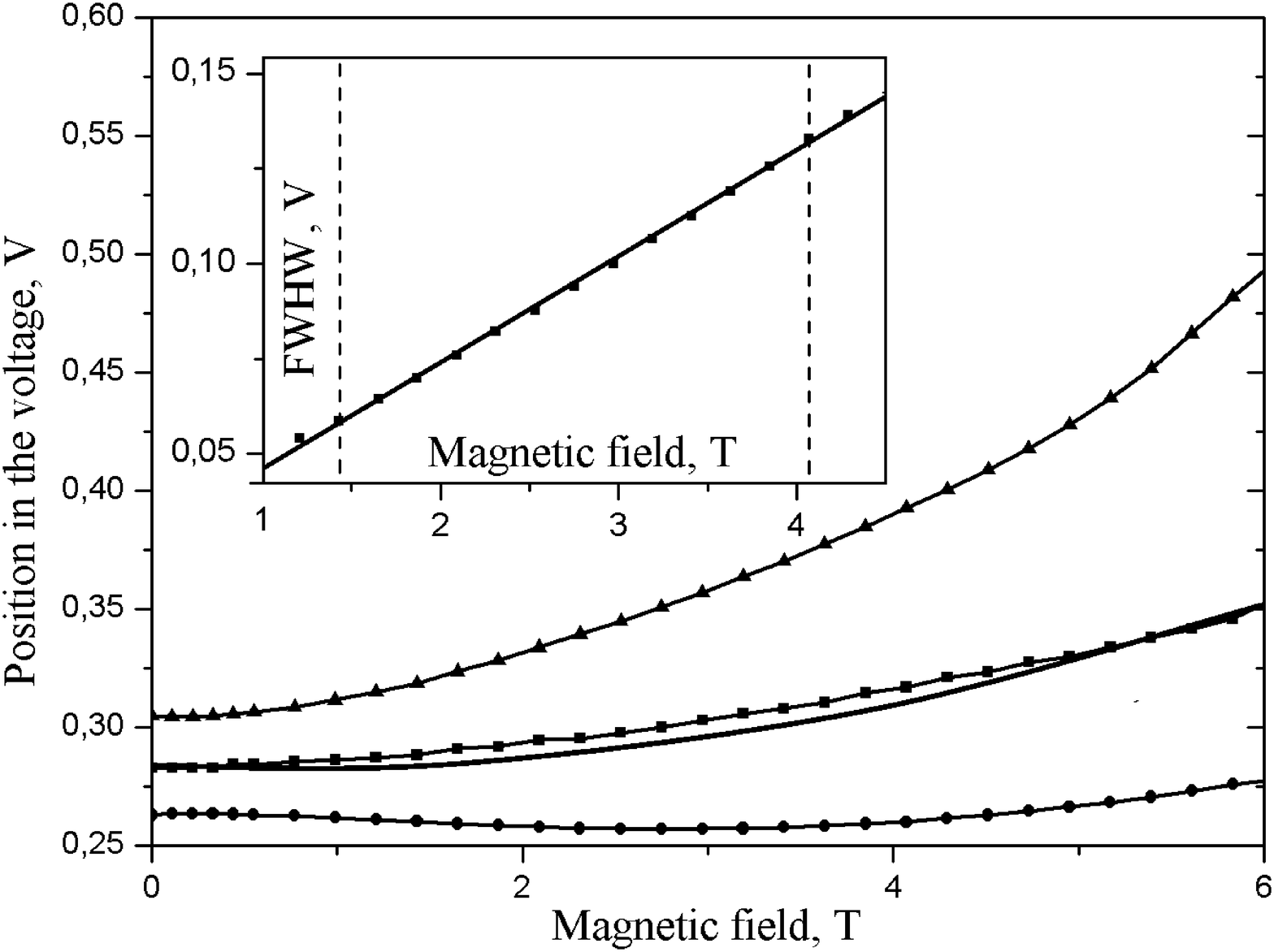}
\caption{\label{fig3} Voltage positions of the the main resonance and the left $U_b$ and  the right $U_e$ its boundaries in the dependence of magnetic field are shown by curves with symbols. The solid curve is a best-fitted parabolla to the experimental positions. Dependence of the full width on half weight of main resonance is shown in the insert symbols. The solid line in the inset is a best-fit of the experimental data. The vertical dotted lines denote the area in which the conditions (7) are met.}
\end{figure}
We observe anticrossing in Fig.\ref{fig2}. It is clearly seen that the main peak amplitude decreases with displacement of the main peak in the phonon replica region and replica amplitude increases. The phenomenon of anticrossing in phonon replicas indicates the presence of polaron states. Anticrossing is described in more detail in Ref. \onlinecite{Polarons2010}.
We approximate the experimental data $U_{peak}(B)$ by the parabolic function $F=A\cdot B^2 + const$, and we obtain the value $A=0.00188~V/T^2$. Thus we can get the following expression:
\be
\alpha \Delta z^2 = 1440~nm^2
\ee
We estimated $\alpha$ by analyzing value of $W$ that is a full width at the half-weight (FWHW) of the main resonance peak. Similar equation to Eq.(\ref{Eq1}) had been derived in Reference \onlinecite{Smoliner91} for the start of the resonance $U_b$ and end of the resonance $U_e$. According to this we analyze the dependence of $W$ of the main resonance on the magnetic field (see Fig. \ref{fig3}) and estimate $\varepsilon_F$ Fermi energy in the emitter. For the estimation we have to use the region on the curves where
\be
B_W < B < B_c
\ee
Here $B_W$ is a magnetic field corresponding to the width of the energy level, $B_c$ is a field when the main current peak approaches to the LO-phonon replicas and anticrosses them\cite{Polarons2010}.
The physical meaning of this condition is following: the intersection point of the emitter dispersion curve and QW electron dispersion curve is in an interval of parabolic part of the spectrum in the QW that is at the electron energy lower LO-phonon energy. In this field region the dependence of $W$ on magnetic field is linear (see inset on Fig. \ref{fig3}) that results in
\be \label{width}
W = U_b - U_e \propto KB
\ee
where $K =$~0.028~V/T is a linear coefficient of the best-fit line shown in Fig.\ref{fig3}.
 Using the value of the $K$, one can determine $\alpha$ as follows:
\be
\alpha = \left(
\frac {1}{m^*\varepsilon_F}
\frac {m^{*2} K^2}{4 (\Delta z)^2}
\right)^{1/2}
\ee
We have got value $\alpha = 5,7$. This estimation has low accuracy because Eq. \ref{width} is disregarded the finite FWHW at zero field $W_0$. Hence we can state this value coincides with that extracted from the perpendicular-field data. It confirms the stability of $\alpha$ in the wide range of fields and voltages.

\section{\label{Dis} Discussion}
We have analyzed the behavior of characteristics in planar and perpendicular magnetic field and estimated the values of the $\varepsilon_F$ and $\alpha$ in our sample. This allows us to calculate the expected positions of the features in the fine structure of the phonon replica. We used the $\alpha$ value obtained from the experiment in the perpendicular field ($\alpha = 5.1$), because it is more accurate.
In LOPAT model according to the Eq. \ref{EqLOPAT} two additional current maxima (and second-derivative minima) $p_1^{**}$ and $p_2^{**}$ should be observed at voltages 0.46 V and 0.55 V. In the case of RTP model according to the Eq. \ref{EqRTP} additional replicas $p_1^*$ and $p_2^*$ should be observed at lower values of voltage by an amount corresponding to the Fermi energy in the emitter, that is, when the voltage values are 0.42 V and 0.51 V. In Fig.\ref{fig10} one can see sharp minima of the second derivative at voltages $V_{p_1^*} = 0.43$~V and $ V_{p_1^{**}} = 0.46$~V, and a broaden minima at voltages $V_{ p_2^*} = 0.51$~V and $ V_{p_2^{**}} = 0.58$~V. Thus a good agreement with the experiment has been succeeded and fine structure of the phonon replica has been described.
\section{\label{Conc} Conclusion}
In summary, the fine structure of phonon replicas had been revealed and described in current-voltage characteristic of the resonant-tunneling diode made of asymmetric double-barrier heterostructure. The phonon replicas were observed at bias voltage when the charge accumulation in the QW is minimal since the thicker barrier layer is closer to the emitter. In this case the leverage factor and the Fermi energy of the emitter 2D layer have been determined and shown almost independent on the current and bias voltage. This allows us to identify the fine-structure features dealt with LOPAT and RTP models.
This result confirms the correctness of the RTP-model and it confirms the existence of polaron states in low-dimensional semiconductor structures. Secondly, it further shows that both tunneling mechanisms (LOPAT and RTP) are taking place in the RTD.
\section{Acknowledgements}
This work was supported by RFBR grant 13-02-01025-a, and the program CNRS-LNCMI ¹ SG-SC0712.
\bigskip\qquad

\end{document}